\documentclass[aps,prb,superscriptaddress,floatfix,notitlepage,twocolumn]{revtex4-2}
\usepackage[utf8]{inputenc}
\usepackage{amsmath}
\usepackage{graphicx}
\usepackage{bm}
\usepackage{hyperref}
\usepackage{mathtools}
\usepackage{color}
\usepackage{soul}
\hypersetup{
 colorlinks = true,%
 linkcolor = blue,
 citecolor = blue,
 urlcolor = blue
}
\setcitestyle{super}

\begin{document}

\title{Lock-in thermography using diamond quantum sensors}

\author{Kensuke Ogawa}
\email{kensuke.ogawa@phys.s.u-tokyo.ac.jp}
\affiliation{Department of Physics, The University of Tokyo, Bunkyo-ku, Tokyo 113-0033, Japan}

\author{Moeta Tsukamoto}
\affiliation{Department of Physics, The University of Tokyo, Bunkyo-ku, Tokyo 113-0033, Japan}

\author{Kento Sasaki}
\affiliation{Department of Physics, The University of Tokyo, Bunkyo-ku, Tokyo 113-0033, Japan}

\author{Kensuke Kobayashi}
\affiliation{Department of Physics, The University of Tokyo, Bunkyo-ku, Tokyo 113-0033, Japan}
\affiliation{Institute for Physics of Intelligence, The University of Tokyo, Bunkyo-ku, Tokyo 113-0033, Japan}
\affiliation{Trans-scale Quantum Science Institute, The University of Tokyo, Bunkyo-ku, Tokyo 113-0033, Japan}

\date{\today}

\begin{abstract}
Precise measurement of temperature distribution and thermal behavior in microscopic regions is critical in many research fields. We demonstrate lock-in thermography using nitrogen-vacancy centers in diamond nanoparticles. We successfully visualize thermal diffusion in glass coverslip and Teflon with micrometer resolution and deduce their thermal diffusivity. By spreading diamond nanoparticles over the sample surface, temperature variation can be measured directly without any physical contact, such as lead wires, making it possible to visualize the micrometer-scale thermal behavior of various materials.
\end{abstract}

\maketitle

\clearpage
\section{Introduction}
Thermometry on a microscopic scale has gained increasing interest in many areas. For example, in condensed matter physics, measuring thermal transport on a microscopic scale is necessary for revealing the behavior of phonons. \cite{phonon,Hydro,Machida} In nanoelectronics, as semiconductor integrated circuits have become miniaturized, \cite{FET} probing local temperature is indispensable for designing efficient heat removal methods. In biosciences, microscopic measurement of intracellular temperature is helpful in understanding cellular and molecular activities. \cite{Nanobio} Thus, developments of thermometers with high spatial resolution are demanded. \par
There are various types of thermometry. Besides usual methods of measuring steady-state conditions, a lock-in method that targets periodic data \cite{Lockin} is powerful for obtaining dynamics, such as the thermal diffusivity of materials, from the amplitude and phase of the temperature oscillation. Additionally, the lock-in method is attractive because of its high sensitivity due to its phase-sensitive nature. A variety of thermometers can be used for this implementation. \cite{Thermometry} Typical examples include thermocouples and resistance thermometers, which are contact-type sensors, infrared radiation thermometers (IR), thermoreflectance, and Raman spectroscopy, which use optics, nano-emitters, which takes advantage of the temperature dependence of the emission spectrum of quantum dots or organic-dye. \cite{nanoemitter1,nanoemitter2,nanoemitter3,nanoemitter4,nanoemitter5} Scanning thermal microscopy (SThM) is also used to obtain high spatial resolution. \cite{SThM} However, each thermometer has its drawbacks: The contact sensors are inevitably invasive, the optical ones are susceptible to the surface properties of the material of interest, nano-emitters often cause photobleaching or photoblinking, and SThM is time-consuming. \par
Nitrogen-vacancy (NV) center is a point defect consisting of a vacancy adjacent to a nitrogen atom in diamond. With its excellent spin properties even at room temperature, \cite{NV} it is promising as a quantum sensor that can probe various physical quantities such as magnetic field, temperature, electric field, and pressure. Since temperature dependence of the zero-field splitting of NV center was pointed out in 2010, \cite{Acosta} NV-center-based thermometry has been shown to have high sensitivity \cite{Wrachtrup,Awschalom} and high spatial resolution. \cite{Kucsko} In most cases, NV centers in nanodiamonds with a particle size of less than $100 ~ \mathrm{nm}$ are used. They work at any temperature above $150 ~ \mathrm{K}$ \cite{chen2011temperature} and at relatively low magnetic fields ($< 1 ~ \mathrm{mT}$). \cite{foy2020wide} The use of a single nanodiamond can, in principle, improve the spatial resolution down to its grain size. To date, nanodiamond thermometry has been studied mainly for physical and biological applications. \cite{Kucsko, Laraoui, Sentyu, foy2020wide} It is also noteworthy that NV centers are suitable for time-resolved thermometry because diamond is one of the most thermally conductive materials at room temperature. \par
In previous studies, time-resolved thermometry using NV centers has been performed \cite{Sentyu,Singam,Tzeng2015} but has never been extended to lock-in measurements. Lock-in thermometry with nanodiamonds should allow us to quantitatively measure local physical quantities such as thermal diffusivity with nanometer-scale spatial resolution. Since thermometry using NV center is realized by spreading nanodiamonds on a sample, it can be applied to various arbitrary-shaped materials, including insulators, semiconductors, and metals. \par
In this paper, we describe lock-in thermography using nanodiamonds and its application to measuring the thermal diffusivity of glass coverslip and Teflon (PTFE, polytetrafluoroethylene) at a micrometer scale as a proof of principle. Here, we define lock-in measurement as an alternating and phase-sensitive measurement. The demonstrated technique has the potential to visualize thermal dynamics down to nanodiamond particle scale. \\

\begin{figure}
    \centering
    \includegraphics[width=\linewidth]{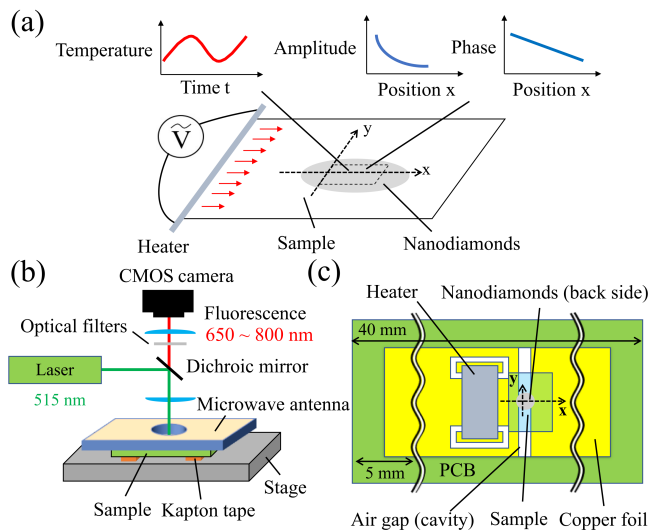}
    \caption{(Color online) (a) Conceptual diagram of our lock-in thermography. In applying the AC heat flux, the periodic temperature variations are measured in a time- and position-resolved manner. The spatial distribution of the phase and amplitude provides information about the thermophysical properties of the target sample. (b) Schematic of the imaging setup using NV centers in nanodiamonds. (c) Setup for the measurement of the heat diffusion on a sample.}
    \label{fig:1}
\end{figure}

\section{Principle}
Figure~1(a) shows the basic idea of the present study. Heat is input from the heater at the left end ($x = x_{h} < 0$). On the assumption that that one-dimensional heat conduction and no heat generation from inside the sample, the heat propagates according to the following heat diffusion equation based on Fourier's law, \cite{Lockin} \\
\begin{equation}
  c_{p} \rho \frac{\partial T(x,t)}{\partial t} = \lambda \frac{\partial^{2} T(x,t)}{\partial x^{2}}.
\end{equation}
Here, $T(x,t)$ is the temperature at a given position $x$ and time $t$, $c_{p}$ is the specific heat, $\rho$ is the density, and $\lambda$ is the thermal conductivity.
Consider a situation where an AC heat flux of angular frequency $\omega$ is input from the heater. Assuming that temperature coincides with room temperature $T_{\textrm{R}}$ at $x = \infty$, the solution to Eq. (1) is as follows. 
\begin{equation}
\begin{split}
  T(x,t) = \frac{p_{0}}{\sqrt{\lambda \rho c_{p} \omega}} \exp \left(-\frac{x-x_{h}}{x_{d}}\right) \times \\ \sin \left(\omega t - \frac{x-x_{h}}{x_{d}} + \phi - \frac{\pi}{4} \right) + T_{\mathrm{R}}.
\end{split}
\end{equation}
Here, $p_{0}$ and $\phi$ are the amplitude and initial phase of the AC heat flux from the heater, respectively. $x_{d} = \sqrt{2\alpha/\omega}$ and $\alpha = \lambda/(\rho c_{p})$, which are the thermal diffusion length and thermal diffusivity, respectively. \par
Equation (2) indicates that the amplitude of the temperature oscillation decays exponentially, and the phase evolves linearly with distance from the heater position. Specifically, the amplitude decay and phase evolution per unit distance are proportional to $x_d$ and $x_d^{-1}$, respectively. Lock-in thermography can estimate the sample's thermal diffusivity from these values. \\

\section{Experimental Setup}
The imaging setup used in this study is schematically shown in Fig. 1(b). \cite{Tsukamoto} A laser beam of $150 ~ \mathrm{mW}$ output power and $515 ~ \mathrm{nm}$ wavelength is expanded by lenses and irradiated onto nanodiamonds dispersed on a sample surface through an objective lens with a magnification of 100 and $\mathrm{NA} = 0.7$. The microwave signal at a frequency centered around $2.87 ~ \mathrm{GHz}$ is output with $-30 ~ \mathrm{dBm}$ from the signal generator, amplified by $45 ~ \mathrm{dB}$, and input to the microwave antenna. \cite{Microwave} The microwave with $33 ~ \mathrm{mW}$ power is uniformly irradiated over a wide area through a circular cavity of $1 ~ \mathrm{mm}$ diameter of the antenna, which means the microwave heating should be negligible.
The fluorescence from the NV centers in nanodiamonds passes through a dichroic mirror, a 514 nm notch filter, a 650 nm long-pass filter, and an 800 nm short-pass filter, and finally, is focused on a CMOS camera. The CMOS camera has $772 \times 1032$ pixels and a field of view of $106 ~ \mathrm{\mu m} \times 140 ~ \mathrm{\mu m}$. The signal-to-noise ratio (SNR) per pixel is improved by integrating the data over multiple pixels. \par
The thermal diffusion is measured using a setup shown in Fig.~1(c). We fabricated a commercial printed circuit board (PCB: FR-4.0) with a copper foil thickness of $18 ~ \mathrm{\mu m}$. A $10 ~ \Omega$ resistor (RK73HW3ATTE10R0F, KOA Corporation, size = $6.3\times3.1\times0.6 ~ \mathrm{mm}^3$) placed on the board is used as a heater by applying a voltage. The thermal contact between the chip resistor and the copper foil is ensured by the thermal grease. The sample is placed in a cavity in the center of the PCB board, and heat is input from the heater through the copper foil. The left edge of the field of view ($x = 0$) is located $100 ~ \mathrm{\mu m}$ away from the left edge of the air gap. The copper foil on the other side serves as a heat sink. All the experiments are conducted at ambient temperature and pressure.\par 
The samples used in the thermal measurement are glass coverslip (thickness:$\  150 ~ \mathrm{\mu m}$, Matsunami Glass Ind., Ltd.) and Teflon (thickness:$\ 200 ~ \mathrm{\mu m}$, The Nilaco Corporation). The nanodiamonds (particle size $50 ~ \mathrm{nm}$, NDNV50nmHi10ml, Adamas Nanotechnologies) are spread on the samples by spin coating at $1000 ~ \mathrm{RPM}$ until the solution is dehydrated. The thickness of the spread nanodiamonds is estimated at $(212 \pm 77) ~ \textrm{nm}$ by the profilometer (KLA Tencor), indicating that the nanodiamonds are piled up about $(4 \pm 1.5)$ layers on the coverslip. \par

\begin{figure}
  \centering
  \includegraphics[width=\linewidth]{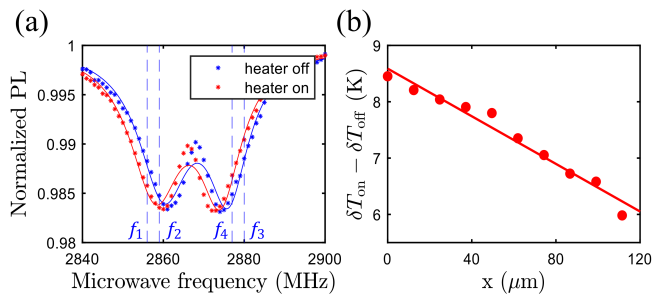}
  \caption{(Color online) (a) ODMR spectra of nanodiamonds on the coverslip with and without DC heating. (b) Temperature gradient on the coverslip due to Joule heating.}
  \label{fig:2}
\end{figure}

\section{Results}
\subsection*{Nanodiamond Thermometry with DC Measurement}
Before performing the lock-in thermography, we verify the operation of our setup [Fig.~1(c)] by measuring the spatial temperature distribution when the heater is energized with a constant input voltage.  We call this DC measurement in DC heating condition. Figure~2(a) shows the optically detected magnetic resonance (ODMR) spectra of the nanodiamonds on the coverslip when the heater is off (blue) and when it is energized with a DC voltage of about $3 ~ \mathrm{V}$ (Joule heat is about 1 W) (red). The horizontal axis represents the microwave frequency, and the vertical axis represents photoluminescence (PL) normalized to that without the application of microwaves. For analysis, the data from all the pixels are summed up. The spectra show that there are two dips around $2860 ~ \mathrm{MHz}$ and $2875 ~ \mathrm{MHz}$. The splitting of about $15 ~ \mathrm{MHz}$ is caused by crystal strain and is a typical value for nanodiamonds. \cite{Kucsko} The center position of these two dips corresponds to the zero-field splitting of the NV center, and it gets smaller by about $2 ~ \mathrm{MHz}$ when voltage is applied. The local temperature can be determined by the zero-field splitting as its temperature coefficient $\kappa$ is known to be $-74.2 ~ \mathrm{kHz/K}$ at around room temperature. \cite{Acosta} Based on this $\kappa$, Fig.~2(a) indicates that the temperature rise of the sample due to the application of voltage is about $27 ~ \mathrm{K}$. \par
To perform imaging, we measure the zero-field splitting for each pixel and convert it to a temperature. To do this efficiently, we use the four-point measurement method. \cite{Kucsko} Instead of acquiring the entire ODMR spectrum, we use only the four characteristic microwave frequencies. For example, the frequencies used when the heater is off are shown as blue dashed lines in Fig.~2(a). The microwave frequencies $f_{1}, f_{2}, f_{3}$, and $f_{4}$ are selected on the slope of the ODMR spectrum to satisfy $\delta f = \frac{f_{2}-f_{1}}{2} = \frac{f_{3}-f_{4}}{2}$ and measure the PL contrast, i.e., the decrease of normalized PL, $C_{1}$, $C_{2}$, $C_{3}$, and $C_{4}$ at each microwave frequency. By approximating the slopes of the spectrum to straight lines, the temperature $T$ can be obtained as follows, \cite{Kucsko}
\begin{eqnarray}  
  \delta T(x,t) &=&  T(x,t) - T_{f_{1},f_{2},f_{3},f_{4}} \nonumber \\
           &=&  \frac{\delta f}{\kappa} \frac{C_{1} + C_{2} - C_{3} - C_{4}}{C_{1} - C_{2} + C_{3} - C_{4}}.
\end{eqnarray}
Here, $T_{f_{1},f_{2},f_{3},f_{4}}$ is the reference temperature where $C_1 = C_3$ and $C_2 = C_4$ are satisfied and $\delta T$ is the deviation from $T_{f_{1},f_{2},f_{3},f_{4}}$. This protocol is one of the NV center-based temperature sensing protocols that can eliminate the effect of static magnetic field. \cite{Kucsko,wojciechowski2018precision,moreva2020practical} \par
We measure the one-dimensional temperature distribution in the field of view for a coverslip sample under DC heating. In DC heating, there is not only Joule heating but also heating on the order of several K by irradiated laser. Assuming that the temperature rise due to the heater and the laser is linearly superimposed, the position dependence of the temperature rise of the sample only due to the heater can be derived by taking the difference between the temperature distribution without DC heating $\delta T_{\textrm{off}}$ and the temperature distribution with DC heating $\delta T_{\textrm{on}}$ (see Appendix A). \par
We sum up the data of pixels in the $y$-axis direction with equal horizontal distances from the heater. For the $x$-axis direction, the data for 100 pixels ($13 ~ \mathrm{\mu m}$) is integrated and processed as a single data point. Figure~2(b) shows the result, which tells that the temperature varies linearly to $x$ as expected from Eq.~(1). The thermal conductivity and heat flow from the heater determine this temperature gradient. Since the heat flow depends sensitively on the thermal contact between the heater and the sample, it is challenging to estimate the heat conductivity precisely from this result. Additionally, it is known that the temperature coefficient can vary from one nanodiamond to another, \cite{foy2020wide} so a separate calibration is required to estimate the absolute temperature more precisely. The solution to this problem is the lock-in thermography described in the next section.

\begin{figure}
  \centering
  \includegraphics[width=\linewidth]{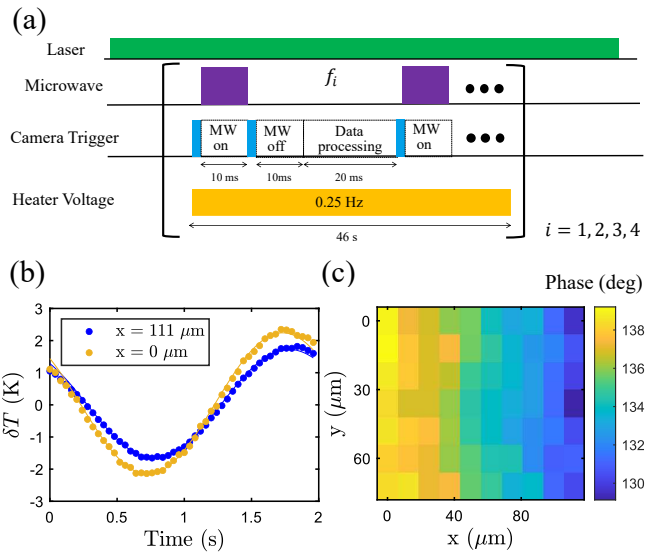}
  \caption{(Color online) (a) Protocol for lock-in measurement. (b) Time-resolved measurement of temperature within a single heating cycle.
  (c) Mapping of the phase of temperature oscillation in the $xy$ plane.}
  \label{fig:3}
\end{figure}

\begin{figure}
  \centering
  \includegraphics[width=\linewidth]{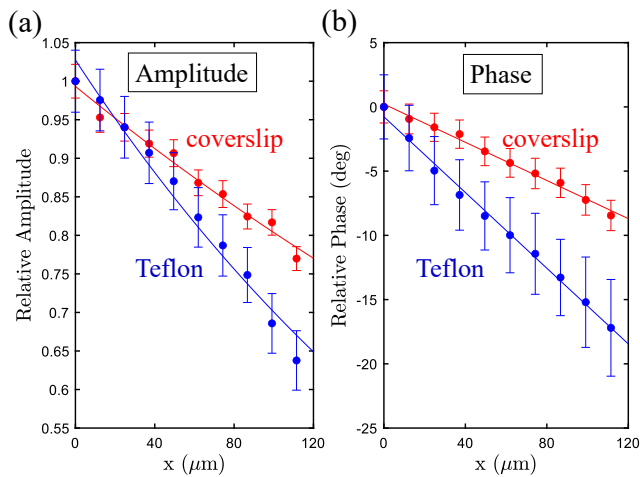}
  \caption{(Color online) (a) Position dependence of the amplitude of temperature oscillation (red: coverslip, blue: Teflon). The amplitude is normalized by the reference value at $x = 0$.
  (b) Position dependence of the phase of temperature oscillation (red: glass coverslip, blue: Teflon). In this study, the phase is subtracted by the reference value at $x = 0$ to focus on the phase change. The error bars correspond to a 95 \% confidence interval of the temperature amplitude and phase.}
  \label{fig:4}
\end{figure}

\subsection*{Lock-in Thermography}
As the validity of our setup has been confirmed, the lock-in measurement is performed. The protocol is shown in Fig.~3(a). An AC voltage of amplitude $3.2 ~ \mathrm{V}$ and frequency $0.25 ~ \mathrm{Hz}$ is applied to the heater to generate thermal oscillations of $0.5 ~ \mathrm{Hz}$, and the temperature distribution is sampled with a time resolution of $40 ~ \mathrm{ms}$. The data is acquired separately for each of the four microwave frequencies. The exposure time is set to $10 ~ \mathrm{ms}$, and the images with and without microwave irradiation are acquired alternately to obtain the PL contrast at each time. A margin of $20 ~ \mathrm{ms}$ is provided for every two acquisitions for data transfer and processing. The microwave frequency is switched every 1150 data points ($46 ~ \mathrm{sec}$). Whenever the frequency is changed, the phase of the AC signal applied to the heating signal is initialized. To allow the system to become stabilized under the AC heating condition, the first 150 data points ($6 ~ \mathrm{sec}$) are discarded each time the heater is initialized. The temperature modulation $\delta T$ is calculated from the PL contrast of the four frequencies at each time point using Eq.~(3). \par
First, we show the results for the glass coverslip. The time-resolved temperature change is shown in Fig.~3(b). The yellow line shows the results in the leftmost region of the sample ($x = 0 ~ \mu \mathrm{m}$), and the blue line shows the results in the rightmost region ($x = 111 ~ \mu \mathrm{m}$) [see Figs.~1(a) and 1(c)]. As shown in the graph, the temperature oscillates at the expected frequency, the amplitude decays, and the phase evolves when it moves away from the heater. Figure~3(c) shows the $xy$-imaging of the phase obtained by fitting the temperature change with a sinusoidal curve. Clearly, the phase lags with distance from the heater, indicating that the heat is diffusing in one dimension. This slight phase lag is reasonable compared to the $547 ~ \mu \mathrm{m}$ thermal diffusion length expected from the literature value. \cite{bunkenti} A more quantitative comparison with literature values will be discussed later. For the region of $100 \times 100$ pixels $(13 ~ \mathrm{\mu m})^2$, the temperature sensitivity is estimated to be $(297 \pm 53) \ \mathrm{mK/\sqrt{Hz}}$ (see Appendix B).   \par
To be more quantitative, the data in the $y$-axis direction is summed up and analyzed in one dimension. The position dependence of the amplitude is shown in red in Fig.~4(a). By fitting relying on Eq.~(2), the thermal diffusivity is obtained to be $(3.5\pm 1.5) \times 10^{-7} ~ \mathrm{{m}^{2} / s}$,  which is consistent with the literature value of coverslip, $4.7 \times {10}^{-7} ~ \mathrm{{m}^{2} / s}$. \cite{bunkenti} We also plot the variation of the phase in one dimension in Fig.~4(b) in red. The phase changes linearly, being consistent with Eq.~(2). The thermal diffusivity is, on the other hand, obtained from the fitting value as $(9.3\pm 1.3) \times {10}^{-7} ~ \mathrm{{m}^{2} / s}$, about twice as large as the value obtained from the amplitude. The reason for this discrepancy will be discussed later. \par
Second, we apply the same protocol to Teflon to show the applicability to different materials. The results of amplitude and phase are shown in Figs.~4(a) and (b) in blue, respectively. The phase delay and the amplitude decay are larger than those of the coverslip, reflecting that the thermal diffusivity of Teflon is smaller than that of the coverslip. The thermal diffusivity obtained by fitting the amplitude is $(1.1\pm 0.57) \times 10^{-7} ~ \mathrm{{m}^{2} / s}$, and that obtained from the phase is $(2.4 \pm 0.3) \times 10^{-7} ~ \mathrm{{m}^{2} / s}$. The literature value for the thermal diffusivity of Teflon is $1.2 \times {10}^{-7} ~ \mathrm{{m}^{2} / s}$, \cite{Teflon} which agrees with the value deduced from the amplitude. Again, the value obtained from the phase is larger by a factor of two, which we discuss in the next section. \\

\section{Discussion}
In the above experiments, we have proven that lock-in thermography agrees with Eq.~(2) and that thermal diffusion on the micron scale is visualized. The thermal diffusivity derived from the amplitude quantitatively agrees with the literature value. This means that the present method is a useful and reliable measurement technique. On the other hand, the thermal diffusivity obtained from the phase was about two times larger. Since both analyses should give the same value, there must be some experimental factors. Three possibilities are discussed in the following. \par
The first is possibly due to the irradiated laser. When the lock-in measurement is performed, the PL count unexpectedly oscillates about $0.05 ~ \%$ with the heating cycle. This means that the intensity of the laser irradiated on the sample oscillates due to heating effects such as thermal expansion and drift of the setup. Recently, an irregular modulation of the ODMR spectrum of nanodiamonds depending on the irradiated laser power near-zero magnetic field has been reported. \cite{Fujiwara} This effect can cause the two dips to behave asymmetrically depending on the laser power, reducing the accuracy of temperature shifts. The same behavior is observed in our nanodiamonds. Since its mechanism is not still understood, it is necessary to investigate the laser-power-dependent ODMR spectrum of the NV center in more detail for precise measurement. Note that heating by laser irradiation is considered to have little effect in this alternative measurement. Most heating effect due to the laser is time-independent. We estimated from nanodiamond measurements that $150 ~ \mathrm{mW}$ laser irradiation produces a temperature distribution of a few K in the field of view. Since this heating is essentially time-independent, its effect is canceled in the AC measurements. Also, during the AC measurement, as mentioned above, the photoluminescence counts from the nanodiamonds oscillate about $0.05 ~ \%$ in synchronization with the thermal cycle. However, assuming that the temperature increase due to laser scales linearly with the laser intensity, the effect of this AC heating is only a few mK. \par 
Secondly, the non-uniformity of the nanodiamond distribution on the surface may affect the result. As we mentioned, the surface profile of the sample obtained by the profilometer indicates a variation of several layers of nanodiamonds in the field of view. Since nanodiamonds themselves have high thermal conductivity, there exist regions where heat conduction through nanodiamonds may partially occur. It is necessary to develop a technique to spread nanodiamonds more evenly or to ensure the uniformity of nanodiamonds. \par
Third, the experimental environment may have room for further improvement. Since all the experiments are conducted at ambient temperature and pressure, heat exchange with the outside can cause an artifact. This can be removed by using a vacuum setup. \par
We suppose that all these factors are responsible for the modulation that cannot be removed in the present lock-in measurements. These factors could have added a complex periodic modulation to the sinusoidal wave of temperature, resulting in a modulation of only the phase. \par
Next, we discuss the limitations of our study. First, absolute temperature cannot be derived in our experiment because the exact temperature dependence of zero-field splitting of nanodiamonds used in our experiments is unknown. However, once the zero-field splitting itself has been calibrated, it is possible to convert the obtained signal to absolute temperature. Second, the derivation of thermal diffusivity based on Eq.~(2) can only be justified with uniform materials. However, it is possible to observe AC temperature signals with this protocol even for materials with disordered complex geometries, and it can be used as a lock-in thermometer. \cite{Lockin} Third, spatial resolution in our experiments is limited because signals from many pixels were integrated to improve SNR. It can be increased up to optical resolution by improvements in the properties of nanodiamonds.

\section{Conclusions} 
In summary, by using the lock-in thermography, we successfully measured the temperature oscillation and determined the thermal diffusivity of the glass coverslip and Teflon on a micrometer scale. The results showed that the thermal diffusivity estimated from amplitude agrees with the literature values. The thermal diffusivity derived from the phase was larger than the literature values for the phase results, for which possible reasons and solutions were discussed. Although the lock-in frequency is currently limited to the frame rate of the CMOS camera (up to 100 fps for our case), combining synchronized measurement and quantum control \cite{SubmHz,ArbFreq,Highreso,meinel2021heterodyne} would allow frequency downconversion of temperature changes in the kHz-GHz band. In the present experiment, the nanodiamonds were spread densely on the sample. However, we expect that measurements beyond the optical resolution are feasible by sparsely spreading the nanodiamonds and accurately estimating the position of nanodiamond particles from the photoluminescence. The demonstrated lock-in thermography using NV centers has the potential to enable temperature measurement and visualization of thermal dynamics of a wide range of materials down to a nanometer scale. \\

\section*{acknowledgement}
We thank MEXT-Nanotechnology Platform Program “Microstructure Analysis Platform” for technical support. M.T. and K.O. acknowledge fnancially support from FoPM, WINGS Program, the University of Tokyo, and Grant-in-Aid for JSPS Fellows. M.T. also acknowledgements fnancial support from Daikin Industries, Ltd. This work was partially supported by the Japan Society for the Promotion of Sscience (Nos. JP19H00656, JP19H05826, JP19H05822, JP20K22325, JP22J21401, JP22J21412, and JP22K03524)

\appendix
\renewcommand{\theequation}{A$\cdot$\arabic{equation}}
\setcounter{equation}{0}
\section*{Appendix A: Calculating one-dimensional temperature gradient during DC heating}
In DC heating, there is laser heating of several K as well as Joule heating. Assuming that the temperature rise due to both is linearly additive, the position dependence of the temperature rise of the sample due to the heater can be calculated by taking the difference in temperature distribution with and without Joule heating. Specifically, first, when Joule heating is absent, the temperature distribution $\delta T_{\textrm{off}}(x)$ obtained using four frequencies $f_{1}, f_{2}, f_{3}$, and $f_{4}$ and Eq.~(3) is expressed as follows using the room temperature $T_{\textrm{R}}$ and the temperature rise due to the laser $\delta T_{\textrm{L}}(x)$, 
\begin{equation}
    \delta T_{\mathrm{off}}(x) = T_{\mathrm{R}} + \delta T_{\mathrm{L}}(x) - T_{f_{1},f_{2},f_{3},f_{4}}.
\end{equation}
Next, in the presence of Joule heating, the temperature distribution $\delta T_{\textrm{on}}(x)$ obtained using the four frequencies $f^{'}_{1}, f^{'}_{2}, f^{'}_{3}$, and $f^{'}_{4}$ is expressed as follows using the temperature rise due to the heater $T_{\textrm{J}}(x)$, 
\begin{equation}
    \delta T_{\mathrm{on}}(x) = T_{\mathrm{R}} + \delta T_{\mathrm{J}}(x) + \delta T_{\mathrm{L}}(x) - T_{f^{'}_{1},f^{'}_{2},f^{'}_{3},f^{'}_{4}}.
\end{equation}
Since the heat from the heater shifts the overall temperature, we change the microwave frequencies used for the protocol to increase sensitivity. By taking this difference, the position dependence of Joule heating can be calculated, although there is an offset due to the different reference temperature. \\

\renewcommand{\theequation}{B$\cdot$\arabic{equation}}
\setcounter{equation}{0}
\renewcommand{\thefigure}{B$\cdot$\arabic{figure}}
\setcounter{figure}{0}

\section*{Appendix B: Sensitivity Estimation}

In our experiment, we consider the $100\times100$ pixels of the camera together as one element $(13~\mathrm{\mu m})^2$ and obtain the image with $N_x=10$ elements in the $x$-direction and $N_y = 7$ elements in the $y$-direction.
Here the $x$-direction is the direction of heat propagation, and ideally, the temperature is independent of the $y$-direction.
We acquire periodic temperature images at 40~ms time intervals during the measurement duration $\tau$.
We divide a cycle of temperature change into $N_t = 50$ elements, and integrate the obtained time series data appropriately.

The precision $\sigma$ in the temperature is estimated as,
\begin{equation}
\sigma^2 = \frac{1}{N_t N_x (N_y-1)}  \sum_{k = 1}^{N_t} \sum_{i = 1}^{N_x} 
\left[ \sum_{j = 1}^{N_y} \{T(x_i,y_j,t_k) - \overline{T}(x_i,t_k)\}^2 \right].
\label{eqs1}
\end{equation}
where $\overline{T}(x_i,t_k)=\frac{1}{N_y}\sum_{j = 1}^{N_y}T(x_i,y_j,t_k)$, and $T(x_i,y_j,t_k)$ is the observed temperature at the position elements $(x_i,y_j)$ and the time element $t_k$.

We use Eq.~(B$\cdot$1) to analyze the $\sigma$ from the temperature data at different $\tau$.
The results are shown in Fig.~B$\cdot$1.
As $\tau$ becomes longer, $\sigma$ decreases, i.e., the accuracy becomes better.
Based on the central limit theorem, we consider $\sigma$ to be inversely proportional to $\sqrt{\tau}$ and fit data with $\sigma=\eta /\sqrt{\tau}$ (solid line).
We then obtain the sensitivity as $\eta = 297 \pm  53 ~ \mathrm{mK}/\sqrt{\mathrm{Hz}}$.

\begin{figure}[htp]
\centering
\includegraphics[width=10cm]{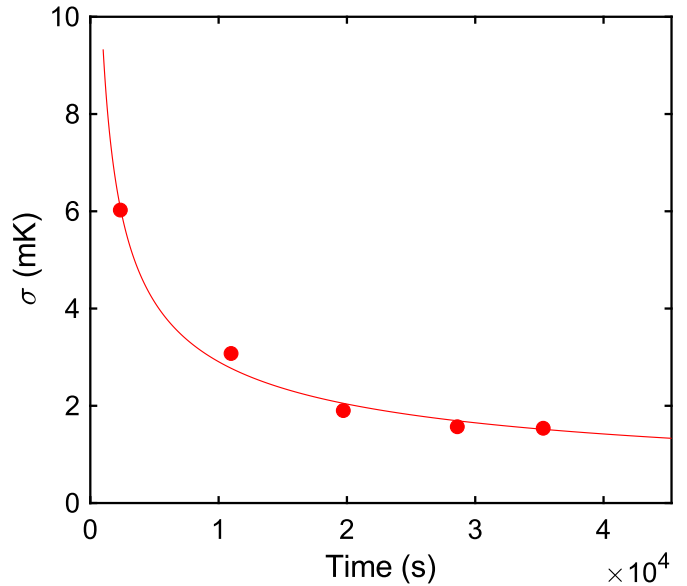}
\caption{Relationship between measurement duration $\tau$ and accuracy $\sigma$.}
\label{sensitivity}
\end{figure}

\clearpage

\end{document}